\documentclass[superscriptaddress,aps,pra,onecolumn,showpacs,nofootinbib,longbibliography,notitlepage]{revtex4-1}
\usepackage{etex}
\usepackage{amsmath,amssymb,amsthm}
\usepackage[colorlinks=true,citecolor=blue,urlcolor=blue]{hyperref}
\usepackage[pdftex]{graphicx}
\usepackage{times,txfonts}
\usepackage{braket}
\usepackage{color}
\usepackage{natbib}
\usepackage{amsmath,blkarray}
\usepackage{mathtools}
\usepackage{latexsym}
\usepackage{tabularx, booktabs}
\usepackage{graphics,epstopdf}
\usepackage{graphicx}
\usepackage{float}
\usepackage{graphicx}
\usepackage{amsfonts}
\usepackage{subcaption}
\usepackage{color,soul}

\newcommand{\be}{\begin{equation}}
\newcommand{\ee}{\end{equation}}
\newcommand{\ba}{\begin{eqnarray}}
\newcommand{\ea}{\end{eqnarray}}

\newcommand{\tr}{\operatorname{Tr}}

\def \hfillx {\hspace*{-\textwidth} \hfill}
\begin{document}
\title{Persistence of quantum violation of macrorealism for large spins even under coarsening of measurement times}

\author{Sumit Mukherjee}
\email{mukherjeesumit93@gmail.com}
\affiliation{Department of Physics, H. N. B. Garhwal Central University, S. R. T. Campus, Badshahi Thaul, Uttarakhand 249 199, India}

\author{Anik Rudra}
\email{rudraanik13@gmail.com}
\affiliation{Department of Physics, H. N. B. Garhwal Central University, S. R. T. Campus, Badshahi Thaul, Uttarakhand 249 199, India}

\author{Debarshi Das}
\email{dasdebarshi90@gmail.com}
\affiliation{Centre for Astroparticle Physics and Space Science (CAPSS), Bose Institute, Block EN, Sector V, Salt Lake, Kolkata 700 091, India}

\author{Shiladitya Mal}
\email{shiladitya.27@gmail.com}
\affiliation{Harish-Chandra Research Institute, HBNI, Chhatnag Road, Jhunsi, Allahabad 211 019, India}

\author{Dipankar Home}
\email{quantumhome80@gmail.com}
\affiliation{Centre for Astroparticle Physics and Space Science (CAPSS), Bose Institute, Block EN, Sector V, Salt Lake, Kolkata 700 091, India}

\begin{abstract}
We investigate quantum violation of macrorealism  for multilevel spin systems under the condition of coarsening of measurement times- i.e., when measurement times have experimental indeterminacy. This is studied together with the effect of coarsening of measurement outcomes for which individual outcomes cannot be unambiguously discriminated. In our treatment, along with different measurement outcomes being clubbed together into two groups in order to model the coarsening of measurement outcomes, importantly, varying degrees of coarsening of measurement time intervals have also been considered. This then reveals that while for a given dimension, the magnitude of quantum violation of macrorealism decreases with the increasing degree of coarsening of measurement times, interestingly, this effect of coarsening of measurement times can be annulled by increasing the dimension of the spin system so that in the limit of large spin, the quantum violation of macrorealism continues to persist. Thus, the result obtained demonstrates that classicality for large spins does not emerge from quantum mechanics in spite of the coarsening of measurement times. 
\end{abstract} 


\maketitle

\section{Introduction}

The key features of quantum mechanics significantly differ from that of classical description of nature. Leggett and Garg, in their seminal paper \cite{lgi1}, codified the central concepts underpinning the classical world view  in terms of the notion of ``macrorealism" which is the conjunction of the following two assumptions: \textit{Realism:} At any instant, irrespective of any measurement, a system is definitely in any one of the available states such that all its observable properties have definite values. \textit{Noninvasive measurability:} It is possible, in principle, to determine which of the states the system is in, without affecting the state itself or the system's subsequent evolution. The conjunction of these two assumptions are in conflict with quantum mechanics which was initially demonstrated by Leggett and Garg by deriving from these assumptions of macrorealism an experimentally testable inequality involving time-separated correlation functions corresponding to successive measurement outcomes pertaining to a system whose state evolves in time. Such an inequality is known as the Leggett-Garg inequality \cite{lgi1,lgi2,lgi3}. In recent years, investigations related to Leggett-Garg inequality have been acquiring considerable significance, as evidenced by a wide range of theoretical and experimental studies \cite{l1,l2,l3,l4,l5,l6,l7,l8,l9,l10,l11,l12,l13,l14,l15,l16,l17,l18,l19,l20,l21,l22,l24,l25,l26,l27,l28,l29,l30,l31}.

 Apart from Leggett-Garg inequalities, two other necessary conditions of macrorealism have been proposed. These are Wigner's form of Leggett-Garg inequalities \cite{saha} and no-signalling in time conditions \cite{nsit}. Wigner's form of Leggett-Garg inequality is derived as a testable consequence of the probabilistic form of macrorealism following the arguments presented in deriving Wigner's form of the local realist inequalities \cite{gwi1, gwi2, gwi3}. On the other hand, no-signalling in time condition is formulated as a statistical version of `noninvasive measurability' to be satisfied by any macrorealist theory. The non-equivalence between these different necessary conditions of macrorealism with respect to the robustness of their quantum violations against unsharp measurement has been established \cite{saha} and also illustrated in a different context \cite{swati1}. Quantum violations of these three necessary conditions of macrorealism contingent upon using most general dichotomic measurements \cite{bum1} as well as two-parameter multi-outcome generalized measurements have also been studied in detail \cite{bum}. Apart from ruling out a class of realist models, quantum violations of any one of the necessary conditions of macrorealism can be invoked to probe non-classicality of the system under consideration. Hence, the quantum violation of macrorealism can be used as a tool for demonstrating quantum-to-classical transition.

One of the key areas in the study of quantum foundations is the investigation of how classicality emerges from quantum mechanics.  Among the various approaches suggested for addressing this issue \cite{book1,book2,book3,book4,book5,book6,book7,book8,book9,b10,b11,b12,b13}, there are the following three strands of prevalent wisdom which are relevant to the present study:

$\bullet$ One view is that classical physics emerges from the predictions of quantum mechanics in the so called `macroscopic' limit when either the system under consideration is of high dimensionality, for example, large spin system, or if a low dimensional system is of large mass, or if it involves large value of any other relevant parameter. 

$\bullet$ The second approach considers that classicality arises from quantum physics under the restriction of coarse-grained measurement (i.e.,  coarsening of measurement outcomes) for which one can empirically resolve only those outcomes of the relevant measurement that are sufficiently well separated \cite{KB,KBB}.

$\bullet$ The recently suggested approach argues that for cases where coarse-grained measurements do not lead to quantum-to-classical transition, coarse-graining of what has been called `measurement references' enables the emergence of classicality  \cite{jk}.

As regards the first approach mentioned above, a number of counterexamples have been pointed out. For instance, in the case of spatial quantum correlations pertaining to spatially separated particles, it has been demonstrated that quantum features in the form of quantum violations of local realist inequalities persist in the `macroscopic' limit, such as for the large number of constituents of the entangled system or for the large dimensions of the constituents of the entangled system \cite{1ap1,1ap2,1ap3,1ap4,1ap5,1ap6,1ap7,1ap8}. On the other hand, in the case of temporal correlations, it has been shown that the quantum violation of Leggett-Garg inequality persists for arbitrary large value of spin of the system under consideration \cite{bdroni,bdroni1,mal}. 

As regards the second approach mentioned above, it has been shown \cite{KB,KBB} that for a class of Hamiltonians governing the time evolution of the system, if one goes into the limit of sufficiently large spins, but can experimentally only resolve outcomes which are well separated, then the measurement statistics and time evolution appear to be consistent with those obtained from classical laws. This suggests that classicality emerges from quantum mechanics under coarse-grained measurements, i.e.,  when the measurement outcomes are coarsened.  However, there are a number of counterexamples, for example, it has been shown that quantum effect in the sense of violating local realism persists under extreme coarse-grained measurements \cite{lrc}. Modeling coarse-grained measurements by clubbing different outcomes into two groups and considering fuzziness of measurements, it has also been shown that quantum violations of macrorealism persist in the large limit of spin \cite{das}. 

Following the third approach mentioned above, it has been demonstrated \cite{jk} that coarsening of `measurement references' can lead to quantum-to-classical transition when it is not enabled by coarsening of measurement outcomes. In particular, it has been shown that while for a given dimension,  the quantum violation of macrorealism decreases under coarsening of `measurement references', this quantum violation can be further decreased by increasing the dimension of the system, thereby reinforcing the emergence of classicality. Recently, the effects of various types of coarsening of `measurement references' on quantum-classical transition have been investigated  for temporal correlations pertaining to two-level systems \cite{cmnew}. In this context it has been shown that coarsening of `measurement references' in terms of  measurement times is the most effective among the different types of `measurement references' involved in probing  quantum-to-classical transition \cite{cmnew}.

Against the above backdrop, considering the quantum violation of macrorealism, we probe the effect of coarsening of measurement outcomes, treated in conjunction with that of measurement references- a study that has yet remained unexplored. Focusing on multilevel spin systems, we first consider projective spin measurements corresponding to multi-outcomes and then model the coarsening of such measurement outcomes by clubbing them into two groups consisting of equal or almost equal number of outcomes. For modeling the coarsening of measurement references, we consider Gaussian coarsening of measurement times.  The central result revealed by the present study is that, for a given dimension, the effect of coarsening of measurement times in showing the emergence  of classicality can be countered by increasing the dimension of the quantum system, thereby illustrating that, for large spin systems, classicality does not necessarily emerge from quantum mechanics under coarsening of measurement times. This is apparently incompatible with the result obtained by Jeong \textit{et al.} \cite{jk}.

We organize this paper in the following way. In Section \ref{sec2} we briefly discuss the three necessary conditions of macrorealism that have been used in our treatment. In Section \ref{sec3}, we explain the measurement context by modeling the coarsening of measurement outcomes as well as that of the measurement times. The analysis of emergence of classicality for multilevel spin systems is presented in Section \ref{sec4} using the three necessary conditions of macrorealism. Finally, in Section \ref{sec5}, while summarizing the key results obtained in this paper, we explain the reason why the central result obtained in this paper differs with that obtained by Jeong \textit{et al.} \cite{jk}.

\section{THE THREE NECESSARY CONDITIONS OF MACROREALISM} \label{sec2}



Based on the two assumptions of macrorealism proposed by Leggett and Garg \cite{lgi1} mentioned earlier, three necessary conditions of macrorealism have been proposed. These are as follows:

\textit{\textbf{Leggett-Garg inequality:}} It was derived by Leggett and Garg \cite{lgi1} as an experimentally testable algebraic consequence of the deterministic form of macrorealism. Let us consider the time evolution of a system consisting of two states, say, $1$ and $2$. Let us define an observable quantity $Q(t)$  which, when measured, can take values $+1$ and $-1$ depending on whether the system is in the state $1$ and $2$, respectively. Next, consider a collection of sets of experimental runs starting from identical initial states each time. In the first set of runs $Q$ is measured at instances, say, $t_1$ and $t_2$, in the second set of runs $Q$ is measured at $t_2$ and $t_3$ and in the third set of experimental runs $Q$ is measured at $t_1$ and $t_3$, where $t_1 < t_2 < t_3$. Based on the assumptions of macrorealism, the following form of Leggett-Garg inequality is obtained:
\begin{equation}
\label{lgi}
K_{LGI}= C_{12} +C_{23} - C_{13} \leq 1,
\end{equation}
where $C_{ij}$=$\langle Q_i Q_j \rangle$ is the two time correlation function of $Q$ measured at instances $t_i$ and $t_j$. Note that the left hand side of inequality (\ref{lgi}) can be evaluated experimentally. Inequality (\ref{lgi}) imposes macrorealist constraint on the temporal correlations pertaining to any two level system.  The magnitude of quantum violation of Leggett-Garg inequality is denoted by the positive values of $(K_{LGI}-1)$.

\textit{\textbf{Wigner's form of Leggett-Garg inequality:}} Similar to Leggett-Garg inequality, here again consider temporal evolution of a two state system where the available states are, say, $1$ and $2$ and consider measurements of $Q$ at instances $t_1$, $t_2$ and $t_3$ ($t_1 < t_2 < t_3$). For deriving Wigner's form of Leggett-Garg inequality \cite{saha}, the notion of realism is invoked implying the existence of overall joint probabilities $\rho (Q_1, Q_2, Q_3)$ pertaining to different combinations of definite values of outcomes for the relevant measurements. On the other hand, the assumption of noninvasive measurability implies that the observable marginal probabilities can be noninvasively measured without affecting the overall joint probabilities. Then, for example, the observable joint probability $P(Q_{1}-, Q_{2}+)$ of obtaining the outcomes $-1$ and $+1$ for the sequential measurements of $Q$ at the instants $t_1$ and $t_2$, respectively, can be written as
\begin{equation}
\begin{split}
P(Q_{1}-, Q_{2}+) &= \sum_{Q_{3}=\pm 1} \rho(-, +, Q_3) = \rho (-, +, +) + \rho(-, +, -).
\end{split}
\end{equation}
Similarly, one can write the other measurable observable joint probabilities $P(Q_{2}+, Q_{3}+)$ and $P(Q_{1}+, Q_{3}+)$ in terms of overall joint probabilities $\rho (Q_1, Q_2, Q_3)$. Using such expressions we get 
\begin{equation}
P(Q_{1}+, Q_{3}+) + P(Q_{1}-, Q_{2}+) - P(Q_{2}+, Q_{3}+) = \rho (+, -, +) + \rho(-, +, -).
\end{equation}
Then, invoking non-negativity of the overall joint probabilities occurring in the above equation, the following three-term Wigner's form of Leggett-Garg inequality can be obtained
\begin{equation}
\label{wlgi}
K_{WLGI} = P(Q_{2}+, Q_{3}+) - P(Q_{1}-, Q_{2}+) - P(Q_{1}+, Q_{3}+)\leq 0.
\end{equation}
This is a specific expression of Wigner's form of Leggett-Garg inequality which we will be using throughout the paper. Similarly, other expressions of Wigner's form of Leggett-Garg inequality involving any number of two-time joint probabilities can be derived by using various combinations of the observable joint probabilities. The magnitude of quantum mechanical violation of Wigner's form of Leggett-Garg inequality is denoted by the positive values of $K_{WLGI}$.

\textit{\textbf{No-Signalling in Time condition:}} No-signalling in time condition is a set of conditions \cite{nsit} introduced as a statistical consequence of noninvasive measurability condition. The statement of no-signalling in time condition is that the measurement outcome statistics for
any observable at any instant is independent of whether any prior measurement has been carried out. Let us consider a system whose time evolution occurs between two possible states. Probability of obtaining the outcome $+1$ contingent upon performing the measurement of a dichotomic observable $Q$ at an instant, say, $t_3$ without any earlier measurement being performed, is denoted by $P(Q_3 +)$. No-signalling in time condition requires that $P(Q_3 +)$ should remain unchanged even when an earlier measurement is made at $t_2$. Mathematically, no-signalling in time condition can be expressed as the following equation.
\begin{align}
	K_{NSIT} &= P(Q_3 +) - [P(Q_2 +, Q_3 +) + P(Q_2 -, Q_3 +)] = 0
	\label{nsit}
\end{align}
Magnitude of quantum non-satisfaction of no-signalling in time condition is quantified by the nonzero values of $K_{NSIT}$.

Violation of any one of the necessary conditions of macrorealism can be invoked for ruling out a class of realist models. Further, quantum violation of any of the necessary conditions of macrorealism can be used as a tool for revealing quantumness in a context that may apparently seem to entail classical behaviour. The present paper will use this latter feature in order to probe quantum-to-classical transition for multi-level large spin systems, subjected to coarsening of measurement outcomes, in conjunction with coarsening of measurement times.

\section{MULTILEVEL SPIN SYSTEM AND THE MEASUREMENT CONTEXT} \label{sec3}
Consider a spin-$j$ system placed in a uniform magnetic field of magnitude $B_0$ along $x$ direction. The relevant Hamiltonian of the system in the units of $\hbar$ is given by,
\begin{equation}
\label{hh}
H = \Omega J_x
\end{equation}
where $J_x$ is the $x$ component of total spin angular momentum $J$ and $\Omega$ is the angular frequency of precession which is proportional to $B_0$. Now consider the measurements of the $z$ component spin, $J_z$, whose possible outcomes are the eigenvalues of $J_z$ operator. The outcomes of $z$ component of spin are denoted by $m$ which can take values from $-j$ to $+j$ for any spin-$j$ system. 

In quantum mechanics, any two outcomes (say, $m_1$ and $m_2$) of the measurement of $z$ component of spin are associated with orthogonal eigenstates. Now, the notion of `neighboring outcomes' \cite{KB,KBB} arises in the real configuration space where the outcomes are detected by observing the post-measurement state of the pointer. For example, the two eigenvalues $m$ and $m+1$ of the $z$ component of spin observable correspond to neighboring outcomes in a real experimental scenario where it is not always possible to distinguish such neighboring outcomes registered in the detector. In the present study we consider that due to measurement interaction, although the state is projected onto one of the eigenstates of $J_z$ operator, owing to limited resolution of the detector,  the neighboring outcomes cannot be distinguished. Thus, this type of coarsening of measurement outcomes actually encapsulates coarsening of outcomes at the level of detecting these outcomes. Here we consider the case of a detector having extremely low resolution so that the detector can only distinguish two groups of outcomes.

In our model of coarsening of measurement outcomes we club the possible outcomes into two groups in such a way that the number of outcomes in these two groups are equal or almost equal.  For any spin-$j$ system, the number of outcomes of $J_z$ measurement is equal to $2j+1$. When $j$ is half-integer, $2j+1$ is even. On the other hand, $2j+1$ is odd for integer values of $j$. Hence, for systems having half-integer spin values it is possible to club the measurement outcomes into two groups where the numbers of outcomes in the two groups are exactly equal. However, in case of systems having integer values of spin it is not possible to divide the possible outcomes equally.  In our measurement scheme we define the observable quantity $Q$ in the following way: for integer $j$, $Q$ is defined in such a way that $Q=-1$ when $m$ takes any  value from $-j$ to $0$, and $Q=+1$ when $m$ takes any value from $+1$ to $+j$. On the other hand, for half-integer $j$, $Q$ is defined in such a way that $Q=-1$ when $m$ takes any value from $-j$ to $-\frac{1}{2}$, and $Q=+1$ when $m$ takes any value from $+\frac{1}{2}$ to $+j$. In the above grouping scheme, the numbers of measurement outcomes in the two groups defined by $Q=+1$ and $Q=-1$ differ by unity when $j$ is integer. When $j$ is half-integer, the numbers of measurement outcomes in the two groups defined by $Q=+1$ and $Q=-1$ are equal. 

This observable quantity $Q$ is measured at instances $t_{1}$, $t_{2}$ and $t_{3}$ $(t_1<t_2<t_3)$.  Note that this type of grouping scheme  has been used earlier in the treatments of the quantum violation of macrorealism \cite{bdroni,das,bum}. Note that the above measurement  scheme has been critically reexamined in the context of Leggett-Garg inequality by Kumari \textit{et al.} \cite{c6kumaipan}.

To summarize the above discussion, we are considering a measurement scenario in which one cannot distinguish between individual outcomes of $J_z$ measurement, but can determine which of the two groups defined above the obtained outcome belongs to.

Considering the above grouping scheme, any typical joint probability distribution $P(Q_i +, Q_j -)$ can be written as follows
\begin{align}
&P(Q_i +, Q_j -) \nonumber \\
&= 
\begin{dcases}
    \sum_{m_i = +1}^{+j} \sum_{m_j=-j}^{0} P( m_i, m_j),& \text{if  $j$ is integer} \\
    \sum_{m_i = +\frac{1}{2}}^{+j} \sum_{m_j=-j}^{-\frac{1}{2}} P(m_i, m_j),& \text{if  $j$ is half-integer}
\end{dcases},
\label{sum}
\end{align}
where $P(m_i, m_j)$ denotes the joint probability of obtaining the outcomes $m_i$ and $m_j$ when measurements corresponding to the operator $J_z$ are performed at instances $t_i$ and $t_j$ ($t_i < t_j$), respectively.   

When coarsening of the measurement times is not considered, the joint probability $P(m_i, m_j)$ can be evaluated as follows,
\begin{align}
P(m_i, m_j) =& \tr \Big[ \Pi_{m_i} U[\Omega (t_i - t_0)] \rho_i U^{\dagger} [\Omega(t_i - t_0)] \Big]  \tr \Big[ \Pi_{m_j} U[\Omega(t_j - t_i)]  \Pi_{m_i} U^{\dagger}[\Omega(t_j - t_i)] \Big],
\label{eqwithout}
\end{align}
where $\rho_i$ is the initial state of the system at $t = t_0$. $\Pi_{m_i}$ and $\Pi_{m_j}$ are the projectors onto the eigenstates of $J_z$ operators associated with eigenvalues $m_i$ and $m_j$, respectively. $U[\Omega(t_i - t_0)] = \exp[ - i \Omega (t_i - t_0) J_x]$ and $U[\Omega(t_j - t_i)]= \exp[ - i \Omega (t_j - t_i) J_x]$ are the unitary time evolution operators from $t=t_0$ to $t=t_i$ and from $t=t_i$ to $t=t_j$, respectively.

Now, in order to take into account the effect of coarsening of the measurement times, we consider that the measurement time intervals $(t_i - t_j)$ ($t_i < t_j$) are not fixed. We rather consider Gaussian distributions of the measurement time intervals $(t_i - t_j)$ around some fixed values $\overline{(t_i - t_j)}$. In this case,  the joint probability $P(m_i, m_j)$ can be evaluated as follows,
\begin{align}
&P(m_i, m_j) \nonumber\\
 & =  \tr \Big[ \Pi_{m_i} \int_{-\infty}^{\infty} d \theta_1 P_{\Delta} [ \theta_1 - \Omega \overline{(t_i - t_0)}] U(\theta_1) \rho_i U^{\dagger}(\theta_1) \Big]  \tr \Big[ \Pi_{m_j}   \int_{-\infty}^{\infty} d \theta_2 P_{\Delta} [ \theta_2 - \Omega \overline{(t_j - t_i)}] U(\theta_2)  \Pi_{m_i} U^{\dagger}(\theta_2) \Big],
\label{eqwith}
\end{align}
where $P_{\Delta} [ \theta_1 - \Omega \overline{(t_i - t_0)}]$ = $\dfrac{1}{\Delta \sqrt{2 \pi}} \exp \Big[- \dfrac{( \theta_1 - \Omega \overline{(t_i - t_0)})^2}{2 \Delta^2} \Big]$ is the Gaussian kernel centered around $\Omega \overline{(t_i - t_0)}$ with standard deviation $\Delta$, $P_{\Delta} [ \theta_2 - \Omega \overline{(t_j - t_i)}]$ is another Gaussian kernel centered around $\Omega \overline{(t_j - t_i)}$ with standard deviation $\Delta$, $U(\theta_k) = \exp[ - i \theta_k J_x]$ with $k$ $\in$ $\{1, 2\}$. Note that $\Delta$ quantifies the degree of coarsening of measurement times.

Now, to develop our treatment, we consider initialization of our system in each experimental run to be in the state $|-j;j\rangle$ at $t=t_0$, where $|m;j\rangle$ denotes eigenstate of $J_z$ operator corresponding to the eigenvalue $m$. Let us choose $\Omega \overline{(t_1 - t_0)} = \pi$, $\Omega \overline{(t_2 - t_1)} = \frac{\pi}{2}$, $\Omega \overline{(t_3 - t_2)} = \frac{\pi}{2}$ (in radian). For an arbitrary $j$, this choice may not give the maximum quantum violation of the Leggett-Garg inequality, Wigner's form of Leggett-Garg inequality, or no-signalling in time condition in the context of our measurement scheme. However, this choice suffices to give a representative indication of the nature of quantum violation of macrorealism for large $j$ under coarsening of measurement outcomes in conjunction with that of measurement times.

 \section{ANALYSES USING different necessary conditions of macrorealism} \label{sec4}

In this Section we discuss the quantum violations of different necessary conditions of macrorealism  in the context of our measurement scenario and the choice of $\Omega \overline{(t_j - t_i)}$  mentioned earlier. 

Using Eqs.(\ref{eqwithout}) and (\ref{eqwith}) one can obtain the following form of any typical joint probability $P(m_i, m_j)$ when coarsening of the measurement time intervals is not considered,
\begin{align}
P(m_i, m_j) =& \Big|d^j_{m_i, -j} \Big(\Omega \overline{(t_i - t_0)}\Big)\Big|^2 \Big|d^j_{m_j, m_i} \Big(\Omega \overline{(t_j - t_i)}\Big) \Big|^2,
\label{eqwithout2}
\end{align}
where $d^j_{m_j, m_i}(\phi)$ is the Wigner's d-matrix.

Similarly, using Eq.(\ref{eqwith}) one can obtain the following form of the joint probability $P(m_i, m_j)$ when coarsening of the measurement time intervals is considered,
\begin{align}
&P(m_i, m_j) \nonumber\\
 & =  \Bigg(\int_{-\infty}^{\infty} d \theta_1 P_{\Delta} [ \theta_1 - \Omega \overline{(t_i - t_0)}] \Big|d^j_{m_i, m_j}(\theta_1) \Big|^2 \Bigg)  \nonumber\\
& \Bigg( \int_{-\infty}^{\infty} d \theta_2 P_{\Delta} [ \theta_2 - \Omega \overline{(t_j - t_i)}] \Big|d^j_{m_j, m_i}(\theta_2) \Big|^2 \Bigg),
\label{eqwith2}
\end{align}

From Eq.(\ref{eqwithout2}) or Eq.(\ref{eqwith2}) along with Eq.(\ref{sum}) one can calculate any joint probability distribution $P(Q_i \pm, Q_j \pm)$ appearing on the left hand side of Wigner's form of Leggett-Garg inequality (\ref{wlgi}) and no-signalling in time condition (\ref{nsit}). Using the joint probability distributions one can then evaluate the correlations $C_{ij} = \sum_{x =\pm} \sum_{y = \pm} P (Q_i = x, Q_j = y)$ appearing on the left hand side of Leggett-Garg inequality (\ref{lgi}).

Note that in the case of half-integer spin systems, our grouping scheme divides the measurement outcomes equally. But in the case of integer spin systems, the number of measurement outcomes in the two groups differ by unity. That is why we subsequently present the analyses for half-integer and integer spin systems separately.

 \begin{table}[t]
\begin{minipage}{0.5\textwidth}
\centering
\begin{tabular}{ |c|c|c|c|c|c|c|c| } 
\hline
& & & & & & &  \\
\textbf{\textit{$j$}} & $\dfrac{3}{2}$  & $\dfrac{15}{2}$ & $\dfrac{35}{2}$ & $\dfrac{55}{2}$ & $\dfrac{97}{2}$ & $\dfrac{147}{2}$ & $\dfrac{199}{2}$  \\
 & & & & & & &  \\
\hline
\textbf{\textit{Magnitude of}} & & & & & & &  \\
\textbf{\textit{quantum violation}} & & & & & & & \\
\textbf{\textit{of Wigner's form}} & $0.250$ & $0.250$ & $0.250$ & $0.250$ & $0.250$ & $0.250$ & $0.250$\\ 
\textbf{\textit{of Leggett-Garg}} & & & & & & & \\
\textbf{\textit{inequality ($K_{WLGI}$)}} & & & & & & & \\
 \hline
\textbf{\textit{Magnitude of}} & & & & & & & \\
\textbf{\textit{quantum}} & & & & & & & \\
\textbf{\textit{non-satisfaction of}} & $0.500$ & $0.500$ & $0.500$ & $0.500$ & $0.500$ & $0.500$ & $0.500$\\ 
\textbf{\textit{no-signalling in}} & & & & & & & \\
\textbf{\textit{time condition}} & & & & & & & \\
 \hline
\end{tabular}
\caption{Magnitudes of quantum violations of the Wigner's form of Leggett-Garg inequality and magnitudes of quantum non-satisfactions of the no-signalling in time condition for half-integer spin ($j$) systems for our grouping scheme of measurement outcomes and the choice of $\Omega \overline{(t_i - t_j)}$  mentioned in the text. This table shows that  the magnitudes of quantum violation of the Wigner's form of Leggett-Garg inequality and the magnitudes of quantum non-satisfactions of the no-signalling in time condition remain constant with respect to different half-integer values of spin ($j$) when coarsening of measurement times is not taken into account.} \label{tabu1}
\end{minipage}
 \hfillx
\begin{minipage}{0.5\textwidth}
            \centering
\begin{tabular}{|*{7}{c|}}
\hline
 & \multicolumn{3}{|c|}{\textit{\textbf{Magnitude of}}} & \multicolumn{3}{|c|}{\textit{\textbf{Magnitude of}}}\\
  & \multicolumn{3}{|c|}{\textit{\textbf{quantum violation}}} & \multicolumn{3}{|c|}{\textit{\textbf{quantum non-satisfaction}}}\\
$j$  & \multicolumn{3}{|c|}{\textit{\textbf{of Wigner's form}}} & \multicolumn{3}{|c|}{\textit{\textbf{of no-signalling}}}\\
& \multicolumn{3}{|c|}{\textit{\textbf{of Leggett-Garg}}} & \multicolumn{3}{|c|}{\textit{\textbf{in time condition}}}\\
& \multicolumn{3}{|c|}{\textit{\textbf{inequality ($K_{WLGI}$) for}}} & \multicolumn{3}{|c|}{\textit{\textbf{($K_{NSIT}$) for}}}\\
\cline{2-7} 
 & \textit{\textbf{$\Delta = 0.25$}} & \textit{\textbf{$\Delta = 0.55$}} & \textit{\textbf{$\Delta = 0.85$}} & \textit{\textbf{$\Delta = 0.55$}} & \textit{\textbf{$\Delta = 0.70$}} & \textit{\textbf{$\Delta = 0.85$}} \\
 \hline 
 \hline
 & & & & & &  \\
 $\dfrac{3}{2}$ & $0.245$ & $0.181$ & $0.063$ & $0.467$ & $0.433$ & $0.389$ \\
  & & & & & &  \\
$\dfrac{9}{2}$ & $0.249$ & $0.206$ & $0.082$ & $0.486$ & $0.457$ & $0.416$ \\
 & & & & & &  \\
$\dfrac{15}{2}$ & $0.250$ & $0.212$ & $0.088$ & $0.490$ & $0.465$ & $0.423$ \\
 & & & & & &  \\
\hline
\end{tabular}
\caption{Magnitudes of quantum violations of the Wigner's form of Leggett-Garg inequality and magnitudes of quantum non-satisfactions of the no-signalling in time condition for different half-integer values of spin $j$ and different degrees ($\Delta$ in radian) of coarsening of measurement times for our grouping scheme of measurement outcomes and the choice of $\Omega \overline{(t_i - t_j)}$  mentioned in the text. This table shows that, for any fixed value of half-integer spin ($j$), the magnitude of quantum violation of the Wigner's form of Leggett-Garg inequality or the magnitude of quantum non-satisfaction of the no-signalling in time condition decreases with the increasing values of $\Delta$. However, importantly, for any fixed value of $\Delta$,  the magnitude of quantum violation of the Wigner's form of Leggett-Garg inequality orthe  magnitude of quantum non-satisfaction of the no-signalling in time condition increases with the  increasing values of $j$.} \label{tabu2}
\end{minipage}
\end{table}

\subsection{Analysis for half-integer spin systems}

 \begin{figure}[t!]
  \centering
\includegraphics[scale=0.5]{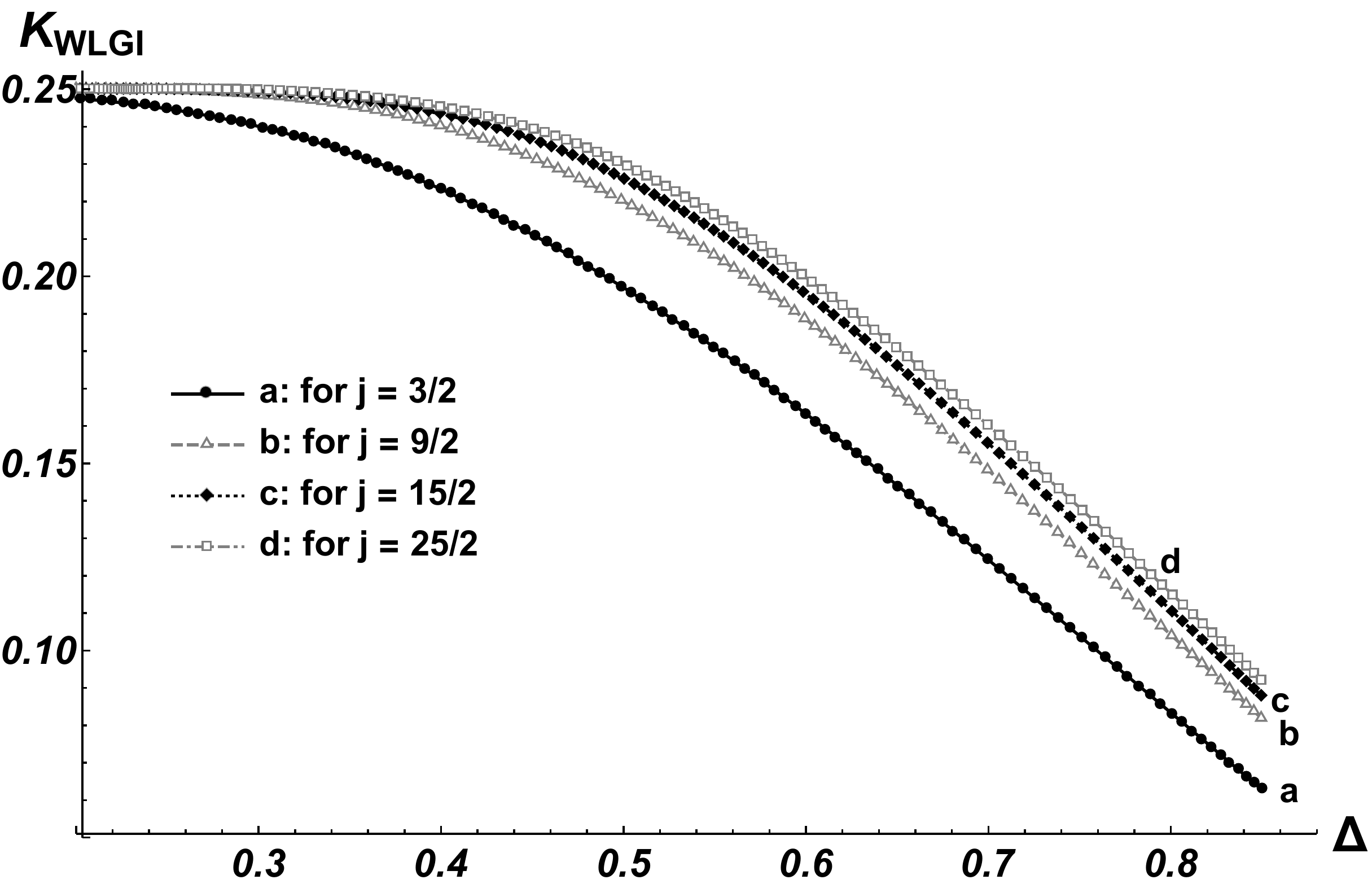}
\caption{Magnitude of quantum violation of the Wigner's form of Leggett-Garg inequality for systems with half-integer values of spin ($j$) under coarsening of measurement times with the degree of coarsening being denoted by $\Delta$ (in radian). From this figure it is evident that, for any fixed value of half-integer spin ($j$), the magnitude of quantum violation of the Wigner's form of Leggett-Garg inequality  decreases with the  increasing values of $\Delta$. On the other hand, importantly, for any fixed value of $\Delta$, the magnitude of quantum violation of the Wigner's form of Leggett-Garg inequality increases with the  increasing values of $j$.}\label{fig1}
\end{figure}

 \begin{figure}[t!]
  \centering
\includegraphics[scale=0.5]{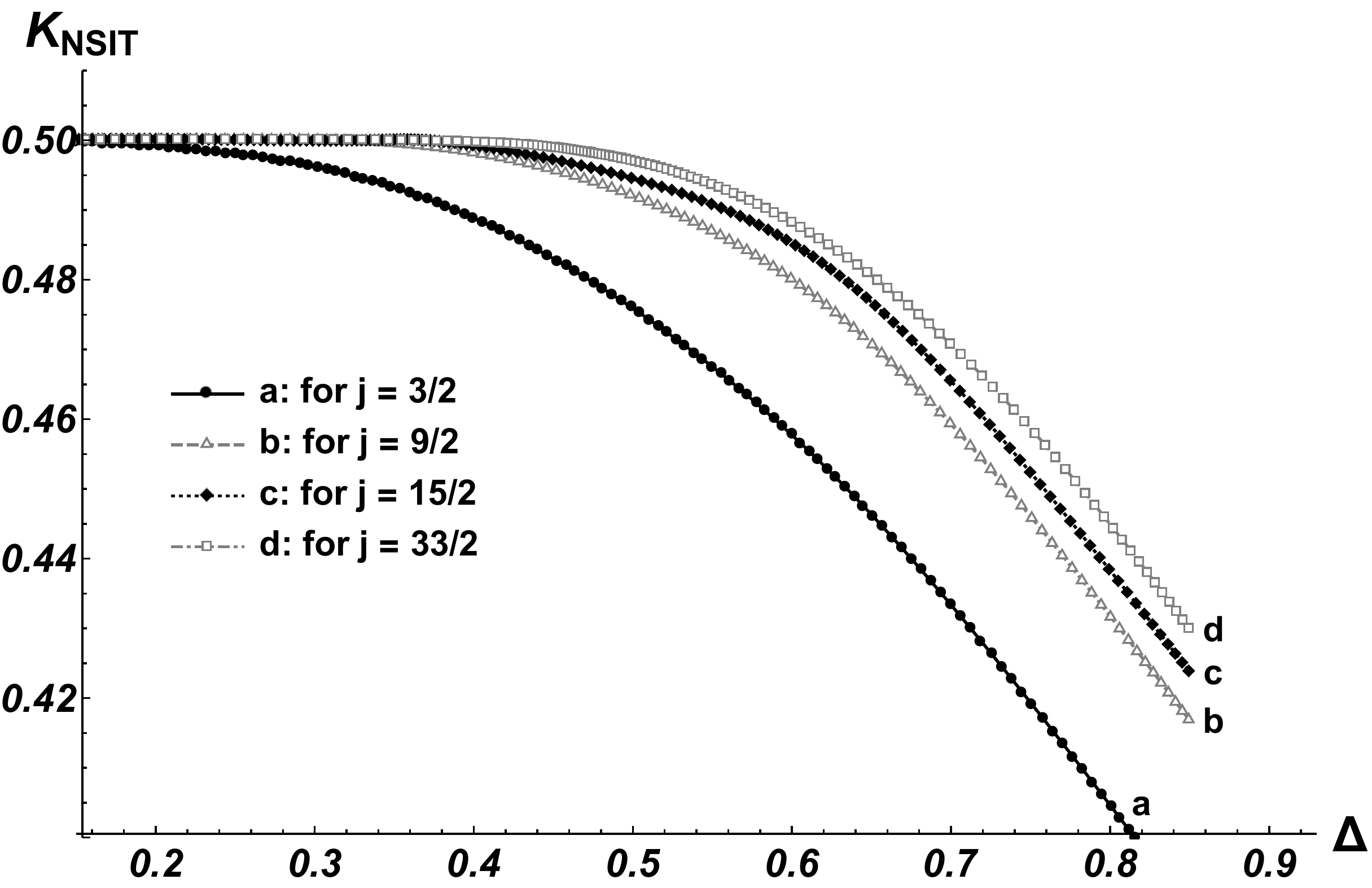}
\caption{Magnitude of quantum non-satisfaction of the no-signalling in time condition for systems with half-integer values of spin ($j$) under coarsening of measurement times with the degree of coarsening being denoted by $\Delta$ (in radian). From this figure it is evident that, for any fixed value of half-integer spin ($j$), the magnitude of  quantum non-satisfaction of the no-signalling in time condition decreases with the  increasing values of $\Delta$. On the other hand, importantly, for any fixed value of $\Delta$, the magnitude of quantum non-satisfaction of the no-signalling in time condition  increases with the  increasing values of $j$.}\label{fig2}
\end{figure}

First, we study the use of Leggett-Garg inequality in our study.  We find that while Leggett-Garg inequality is not violated for arbitrary half-integer spin systems under coarsening of measurement times, even in the absence of coarsening of measurement times Leggett-Garg inequality is not violated. Note that these results are for the particular choice of $\Omega \overline{(t_j - t_i)}$ and particular grouping scheme of the measurement outcomes mentioned earlier.  Since, Leggett-Garg inequality is a necessary condition of macrorealism, but not a sufficient one, no conclusion can be drawn when Leggett-Garg inequality is not violated.

As we will discuss later, when coarsening of measurement times is not considered, Wigner's form of Leggett-Garg inequality is violated and no-signalling in time condition is not satisfied by quantum systems having arbitrary half-integer values of spin for the aforementioned  choice of $\Omega \overline{(t_j - t_i)}$. These quantum effects persist even under coarsening of measurement times. Hence, the particular choice of $\Omega \overline{(t_j - t_i)}$ mentioned earlier suffices to illustrate the main result of this work, namely, the persistence of quantum violation of macrorealism for large spin systems even under coarsening of measurement times, evidenced through quantum violations of the Wigner's form of Leggett-Garg inequality and quantum non-satisfaction of the no-signalling in time condition. 

Next, we study the quantum violations of Wigner's form of Leggett-Garg inequality (\ref{wlgi}). In this case, when coarsening of measurement times is not taken into account, the quantum violation of Wigner's form of Leggett-Garg inequality remains constant with respect to different values of the half-integer spin $j$. This result is summarized in Table \ref{tabu1}.

Now, consider that coarsening of measurement times is taken into account. The result obtained for Wigner's form of Leggett-Garg inequality in this case is summarized in Table \ref{tabu2}. From this Table it is clear that, for any fixed value of half-integer spin, the magnitude of quantum violation of Wigner's form of Leggett-Garg inequality decreases with the  increasing values of $\Delta$. In other words, for a given spin, the magnitude of quantum violation of Wigner's form of Leggett-Garg inequality decreases with the  increasing degree of coarsening of the measurement time intervals. On the other hand, interestingly, for any fixed value of $\Delta$, the magnitude of quantum violation of Wigner's form of Leggett-Garg inequality increases with the  increasing values of $j$. These results are illustrated in Fig. \ref{fig1}. Hence, the effect of coarsening of measurement times in reducing the quantum violation of macrorealism in any given dimension can be countered by increasing the dimension of the system, thereby showing the persistence of quantum violation of macrorealism for large spin systems even under coarsening of measurement times for the grouping scheme of the measurement outcomes considered.

Next, we investigate whether no-signalling in time condition (\ref{nsit}) is satisfied for systems with half-integer values of spin. In this case also, when coarsening of measurement times is not considered, the no-signalling in time condition is not satisfied and the magnitude of quantum non-satisfaction of no-signalling in time condition ($K_{NSIT}$) remains constant with respect to different values of the half-integer spin $j$. This result is summarized in Table \ref{tabu1}. Note that the constant magnitude of quantum non-satisfaction of no-signalling in time condition in this case is greater than the magnitude of quantum violation of Wigner's form of Leggett-Garg inequality mentioned in Table \ref{tabu1}. Hence, for experimental purposes no-signalling in time condition is more effective than Wigner's form of Leggett-Garg inequality for showing non-classicality through the quantum violations of macrorealism.

We now consider that coarsening of measurement times is taken into account. The result obtained for no-signalling in time condition given by Eq.(\ref{nsit}) in this case is summarized in Table \ref{tabu2}. From this Table it is clear that, for any fixed value of half-integer spin, the magnitude of  quantum non-satisfaction of the no-signalling in time condition decreases with the increasing degree of  coarsening of the measurement times. On the other hand, for any fixed value of $\Delta$,  the magnitude of quantum non-satisfaction of no-signalling in time condition  increases with the increasing values of $j$. These results are illustrated in Fig. \ref{fig2}. Hence, similar to the case of the quantum violation of Wigner's form of Leggett-Garg inequality, the effect of coarsening of measurement times in decreasing   the magnitude of quantum non-satisfaction of no-signalling in time condition  in any given dimension can be countered by increasing the dimension of the system. Here it is also important to note that for empirical demonstration of these features, no-signalling in time condition is more effective than Wigner's form of Leggett-Garg inequality since   the magnitude of quantum non-satisfaction of no-signalling in time condition ($K_{NSIT}$)  is much larger than  the magnitude of quantum violation of Wigner's form of Leggett-Garg inequality ($K_{WLGI}$) for any $j$ and $\Delta$, as evidenced by comparing Figs. \ref{fig1} and \ref{fig2}.

\begin{table}[t]
\begin{minipage}{0.5\textwidth}
 \centering
\begin{tabular}{ |c|c|c|c|c|c|c| }  
\hline
\textbf{\textit{$j$}} & $3$ & $9$ & $12$ & $25$ & $40$ & $80$   \\
\hline
\textbf{\textit{Magnitude of}} & & & &  & &  \\
\textbf{\textit{quantum violation}} & & & &  & &  \\
\textbf{\textit{of Wigner's form}} & $0.147$ & $0.195$ & $0.205$ & $0.218$ & $0.226$ & $0.233$ \\ 
\textbf{\textit{of Leggett-Garg}} & & & &  & &  \\
\textbf{\textit{inequality ($K_{WLGI}$)}} & & & &  & &  \\
 \hline
 \textbf{\textit{Magnitude of}} & & & &  & &  \\
\textbf{\textit{quantum}} & & & &  & &  \\
\textbf{\textit{non-satisfaction of}} & $0.451$ & $0.482$ & $0.486$ & $0.493$ & $0.496$ & $0.498$\\ 
\textbf{\textit{no-signalling in}} & & & &  & &  \\
\textbf{\textit{time condition}} & & & &  & &  \\
 \hline
\end{tabular}
\caption{Magnitudes of quantum violations of the Wigner's form of Leggett-Garg inequality and magnitudes of quantum non-satisfactions of the no-signalling in time condition for integer spin ($j$) systems for our grouping scheme of measurement outcomes and the choice of $\Omega \overline{(t_i - t_j)}$  mentioned in the text. This table shows that the magnitudes of quantum violation of the Wigner's form of Leggett-Garg inequality and the magnitudes of quantum non-satisfactions of the no-signalling in time condition increase with increasing integer values of spin ($j$) when coarsening of measurement times is not taken into account.} \label{tabu5}
\end{minipage}
 \hfillx
\begin{minipage}{0.5\textwidth}
            \centering
\begin{tabular}{|*{7}{c|}}
\hline
 & \multicolumn{3}{|c|}{\textit{\textbf{Magnitude of}}} & \multicolumn{3}{|c|}{\textit{\textbf{Magnitude of}}}\\
  & \multicolumn{3}{|c|}{\textit{\textbf{quantum violation}}} & \multicolumn{3}{|c|}{\textit{\textbf{quantum non-satisfaction}}}\\
$j$  & \multicolumn{3}{|c|}{\textit{\textbf{of Wigner's form}}} & \multicolumn{3}{|c|}{\textit{\textbf{of no-signalling}}}\\
& \multicolumn{3}{|c|}{\textit{\textbf{of Leggett-Garg}}} & \multicolumn{3}{|c|}{\textit{\textbf{in time condition}}}\\
& \multicolumn{3}{|c|}{\textit{\textbf{inequality ($K_{WLGI}$) for}}} & \multicolumn{3}{|c|}{\textit{\textbf{($K_{NSIT}$) for}}}\\
\cline{2-7} 
 & \textit{\textbf{$\Delta = 0.25$}} & \textit{\textbf{$\Delta = 0.55$}} & \textit{\textbf{$\Delta = 0.85$}} & \textit{\textbf{$\Delta = 0.55$}} & \textit{\textbf{$\Delta = 0.70$}} & \textit{\textbf{$\Delta = 0.85$}} \\
 \hline 
 \hline
 $3$ & $0.161$ & $0.140$ & $0.053$ & $0.421$ & $0.398$ & $0.365$ \\
$6$ & $0.196$ & $0.176$ & $0.073$ & $0.456$ & $0.433$ & $0.396$ \\
$9$ & $0.210$ & $0.190$ & $0.081$ & $0.468$ & $0.446$ & $0.408$ \\
 $12$ & $0.218$ & $0.198$ & $0.085$ & $0.475$ & $0.453$ & $0.415$ \\
\hline
\end{tabular}
\caption{Magnitudes of quantum violations of the Wigner's form of Leggett-Garg inequality and magnitudes of quantum non-satisfactions of the no-signalling in time condition for different integer values of spin $j$ and different degrees ($\Delta$ in radian) of coarsening of measurement times for our grouping scheme of measurement outcomes and the choice of $\Omega \overline{(t_i - t_j)}$  mentioned in the text. This table shows that, for any fixed value of integer spin ($j$), the magnitude of quantum violation of the Wigner's form of Leggett-Garg inequality or the magnitude of quantum non-satisfaction of the no-signalling in time condition decreases with the increasing values of $\Delta$. However, importantly, for any fixed value of $\Delta$,  the magnitude of quantum violation of the Wigner's form of Leggett-Garg inequality or the magnitude of quantum non-satisfaction of the no-signalling in time condition increases with the  increasing values of $j$.} \label{tabu6}
\end{minipage}
\end{table}

\subsection{Analysis for integer spin systems}

 \begin{figure}[t!]
 \centering
\includegraphics[scale=0.5]{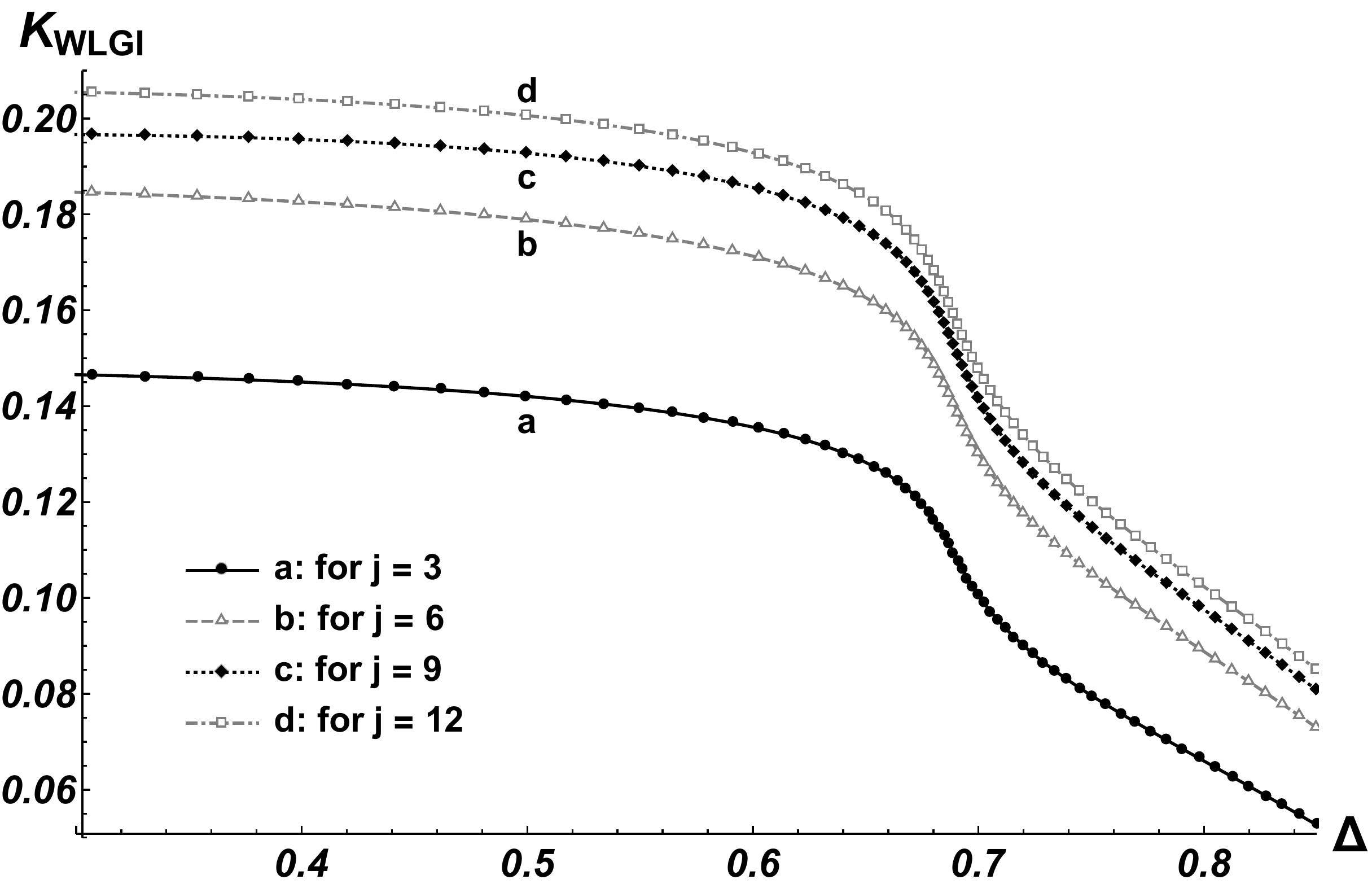}
\caption{Magnitude of quantum violation of the Wigner's form of Leggett-Garg inequality for systems with integer values of spin ($j$) under coarsening of measurement times with the degree of coarsening being denoted by $\Delta$ (in radian). From this figure it is evident that, for any fixed value of integer spin ($j$), the magnitude of quantum violation of the Wigner's form of Leggett-Garg inequality decreases with the  increasing values of $\Delta$. On the other hand, importantly, that for any fixed value of $\Delta$, the magnitude of quantum violation of the Wigner's form of Leggett-Garg inequality increases with the   increasing values of $j$.}\label{fig3}
\end{figure}

\begin{figure}[t]
 \centering
\includegraphics[scale=0.5]{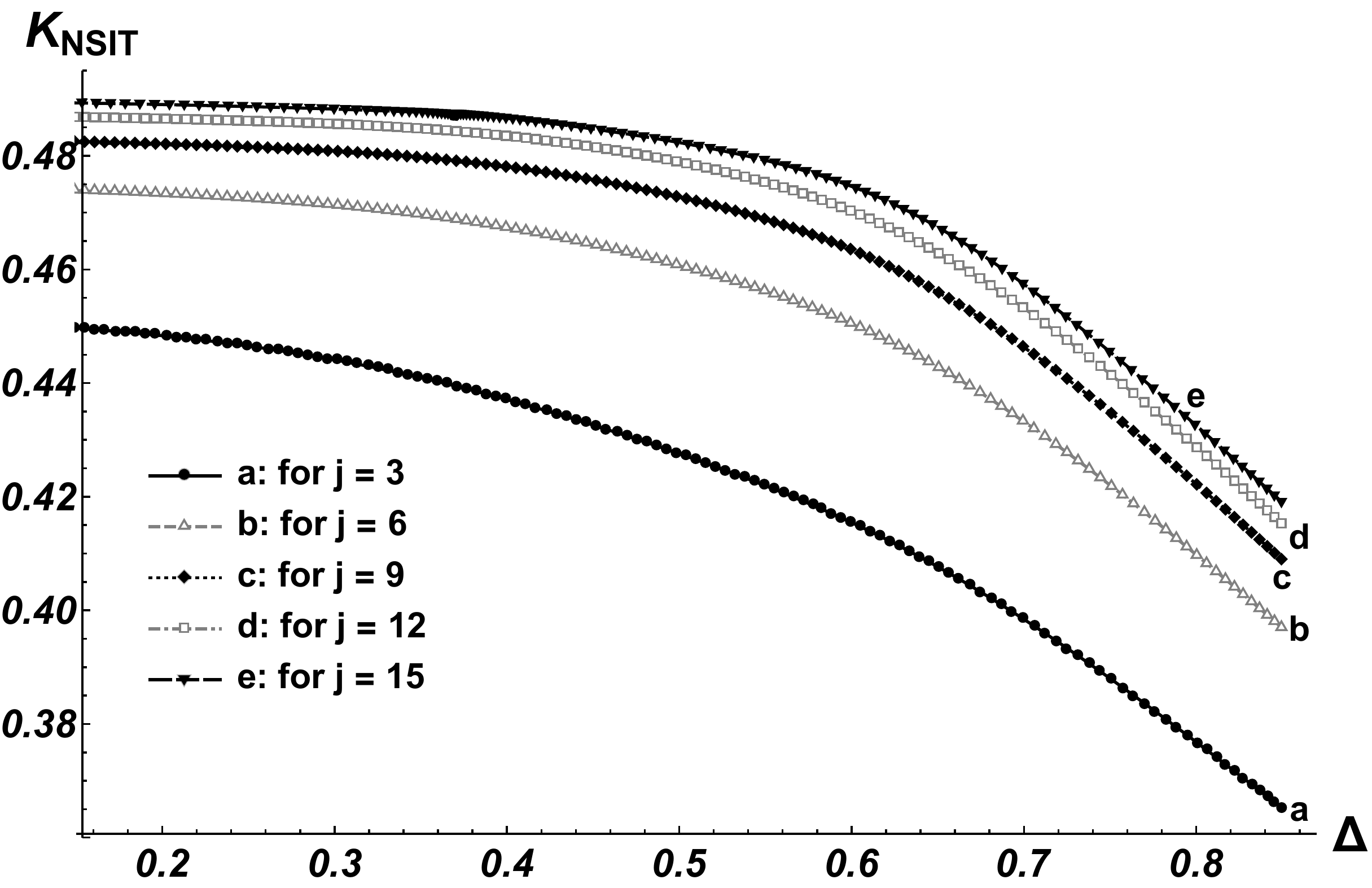}
\caption{Magnitude of quantum non-satisfaction of the no-signalling in time condition in time ($K_{NSIT}$)  for systems with integer values of spin ($j$) under coarsening of measurement times with the degree of coarsening being denoted by $\Delta$ (in radian). From this figure it is evident that, for any fixed value of integer spin ($j$), the magnitude of quantum non-satisfaction of the no-signalling in time condition decreases with the  increasing values of $\Delta$. On the other hand, importantly, for any fixed value of $\Delta$, the magnitude of quantum non-satisfaction of the no-signalling in time condition  increases with the  increasing values of $j$.}\label{fig4}
\end{figure}

In this subsection we consider systems with integer values of spin. In this case also we find that Leggett-Garg inequality  is not violated for arbitrary integer spin systems even when coarsening of measurement times is not taken into account. Therefore, as expected, no quantum violation of Leggett-Garg inequality  is observed when coarsening of measurement times is considered. Note that, as mentioned earlier, this result is valid for the particular choices of $\Omega \overline{(t_j - t_i)}$ and particular grouping scheme of the measurement outcomes.  Similar to the case of half-integer spins, in this case also, quantum violation of macrorealism for large spin systems under coarsening of measurement times is manifested through the Wigner's form of Leggett-Garg inequality (\ref{wlgi}) or through the no-signalling in time condition (\ref{nsit}). This is because Wigner's form of Leggett-Garg inequality (\ref{wlgi}) is violated and the no-signalling in time condition (\ref{nsit}) is not satisfied by systems with arbitrary integer spins under the aforementioned choices of $\Omega \overline{(t_j - t_i)}$ and the particular grouping scheme of the measurement outcomes when coarsening of measurement times is considered. In the following we will elaborate these features.

Now, we present quantum violation of Wigner's form of Leggett-Garg inequality (\ref{wlgi}) for the grouping scheme of the measurement outcomes mentioned earlier. In this case, when coarsening of measurement times is not taken into account, the magnitude of quantum violation of Wigner's form of Leggett-Garg inequality increases with the increasing values of spin $j$. This result is summarized in Table \ref{tabu5}. Note that the magnitude of quantum violation of Wigner's form of Leggett-Garg inequality is not fixed in this case, contrary to the case of half-integer spin systems.

Next, we take into account the effect of coarsening of measurement times considering the aforementioned grouping scheme of the measurement outcomes. The result obtained for Wigner's form of Leggett-Garg inequality in this case is summarized in Table \ref{tabu6}. Similar to the case of half-integer spin systems, in this case also the magnitude of quantum violation of Wigner's form of Leggett-Garg inequality ($K_{WLGI}$) decreases with the increasing degree of coarsening of the measurement time intervals for any fixed value of integer spin. However, the effect of coarsening of measurement times in decreasing quantum effect in the form of quantum violation of Wigner's form of Leggett-Garg inequality in any given dimension is countered by increasing the spin of the system. These results are illustrated in Fig. \ref{fig3}. 

Now, we study the no-signalling in time condition given by  Eq.(\ref{nsit}) for integer spin systems for the grouping scheme of the measurement outcomes we have used. When the coarsening of measurement times is not considered, the no-signalling in time condition (\ref{nsit}) is not satisfied by any integer spin systems and the magnitude of quantum non-satisfaction of no-signalling in time condition increases with the increasing values of spin $j$. This result is summarized in Table \ref{tabu5}. 

Next, we consider that the effect of coarsening of measurement times is taken into account. The results obtained for no-signalling in time condition given by  Eq.(\ref{nsit}) in this case is summarized in Table \ref{tabu6}. From this Table it is evident that  the magnitude of quantum non-satisfaction of  no-signalling in time condition decreases with the increasing degree of coarsening of the measurement time intervals for any fixed integer value of spin. On the other hand, for any fixed value of $\Delta$,  the magnitude of quantum non-satisfaction of no-signalling in time condition increases with the increasing values of $j$. These results are illustrated in Fig. \ref{fig4}. For integer spin systems, too, no-signalling in time  condition is more effective than Wigner's form of Leggett-Garg inequality in experimentally demonstrating all the above features, since  the magnitude of quantum non-satisfaction of no-signalling in time condition is much larger than the magnitude of quantum violation of Wigner's form of Leggett-Garg inequality ($K_{WLGI}$) for any $j$ and $\Delta$ as evident by comparing the Figs. \ref{fig3} and \ref{fig4}.

\textit{Remark:} In the present study we have assumed that $\Omega \overline{(t_1 - t_0)} = \pi$, $\Omega \overline{(t_2 - t_1)} = \Omega \overline{(t_3 - t_2)} = \frac{\pi}{2}$. This fixed choice of mean value of the Gaussian distribution of measurement time intervals does not always give the optimal violations of different necessary conditions of macrorealism for arbitrary values of spin $j$. But this choice enables one to reach the main conclusion of this paper, namely, persistence of quantum violation of macrorealism for large spin under coarsening of measurement times. With the above choice of $\Omega \overline{(t_j - t_i)}$ this conclusion is reached through the Wigner's form of Leggett-Garg inequality and the no-signalling in time  condition. In order to strengthen our result, we have also studied quantum violations of macrorealism in the context mentioned in Section \ref{sec3} with $\Big\{ \Omega \overline{(t_1 - t_0)} = \pi$, $\Omega \overline{(t_2 - t_1)} = \pi$, $\Omega \overline{(t_3 - t_2)} = \pi \Big\} $; $ \Big\{ \Omega \overline{(t_1 - t_0)} = \pi/2$, $\Omega \overline{(t_2 - t_1)} = \pi/2$, $\Omega \overline{(t_3 - t_2)} = \pi \Big\}$; $ \Big\{ \Omega \overline{(t_1 - t_0)} = \pi$, $\Omega \overline{(t_2 - t_1)} = \pi/4$, $\Omega \overline{(t_3 - t_2)} = \pi/4 \Big\}$; $ \Big\{ \Omega \overline{(t_1 - t_0)} = \pi$, $\Omega \overline{(t_2 - t_1)} = 3 \pi/4$, $\Omega \overline{(t_3 - t_2)} = 3 \pi/4 \Big\}$. In these cases we study the effect of coarsening of measurement times on the quantum violation of any necessary condition of macrorealism whenever that necessary condition is violated without considering coarsening of measurement times. With the above choices of $\Omega \overline{(t_i - t_j)}$  we observe that the nature of quantum violation of macrorealism remains unchanged, i. e., the effect of coarsening of measurement times in decreasing quantum violation of macrorealism in any given dimension is countered by increasing the spin of the system. However, for different choices of $\Omega \overline{(t_j - t_i)}$ the above feature is manifested with different necessary conditions of macrorealism. Optimizing quantum violations of different necessary conditions of macrorealism over arbitrary values of $\Omega \overline{(t_j - t_i)}$ merits further investigation which will strengthen the results obtained in this paper.

\section{CONCLUDING DISCUSSION} \label{sec5}

In the present work, we have considered multi-level spin systems and multi-outcome spin measurements. The coarsening of measurement outcomes has been modelled by clubbing different measurement outcomes into two groups consisting of equal or almost equal number of outcomes. We have also considered varying degrees of Gaussian coarsening of measurement time intervals. The interesting result obtained in the present study is that, along with coarsening of the measurement outcomes, the effect of coarsening of measurement times in reducing the magnitude of quantum violation of macrorealism can be compensated by increasing the dimension of the quantum system. Thus, classicality does \textit{not} always emerge from quantum mechanics under coarsening of measurement times for large spin system, contrary to the result obtained in \cite{jk}. For our specific choice of mean measurement time intervals and grouping scheme of measurement outcomes, this feature has been illustrated through the quantum violations of Wigner's form of Leggett-Garg inequality and No-Signalling in Time condition.

As discussed earlier, our adopted model of coarsening of measurement outcomes corresponds to coarsening at the level of detecting the outcomes. While the measuring apparatus can project the state of the system under consideration onto any one of the eigenstate of the spin component observable, the detector accuracy is limited in distinguishing individual outcomes. This has been modelled in our treatment for the case of a detector having very low resolution. In this model, a limitation is that the demarcation between the two groups defined by $Q=+1$ and $Q=-1$ is taken to be precise. In a more realistic coarsening of measurement  outcomes, imprecision involved in this demarcation should also be taken into account. Incorporating this effect in the context of our model of coarsening of measurement outcomes should be worth studying in future.

 In order to pinpoint the reason for the apparent incompatibility of our results with that of Ref. \cite{jk}, we note that Jeong \textit{et al.} considered a $N$ level system and the measurements considered by them project the state onto one of the $2$ possible subspaces and then they considered coarsening of the measurement times. To be precise, the measurements considered in  \cite{jk} correspond to Luder's rule for dichotomic measurements \cite{luder}. On the other hand, the treatment presented in this paper consists of three stages: (a) For a $N$ ($= 2 j+1$) level system, we consider the measurements that project the state of the system onto one of the $N$ subspaces. In other words, we consider the case of a complete degeneracy-breaking measurement \cite{bdroni}, as initially proposed by von-Neumann \cite{vn1,vn2}. (b) Subsequently, for our analysis, in order to take into account the coarsening of individual measurement outcomes, we club the different outcomes into two groups. (c) Then, it is in the context of such grouping scheme of measurement outcomes, we consider coarsening of measurement time intervals. Hence, to summarize, the measuring apparatus associated with the projective measurements used by Jeong \textit{et al.} \cite{jk}, unlike that considered in the present work, do not project the state onto any one of the $(2j+1)$ eigenstates of the spin-$j$ component observable. Hence, the information gain due to the projective measurements used in the present work is more with respect to that used in Ref. \cite{jk}. It is due to this key difference between the measurement scheme used in this paper and that adopted by Jeong \textit{et al.} \cite{jk}, the central result of this paper differs from that obtained in \cite{jk}.


Finally, we may remark that while it is of general fundamental significance to study how classicality emerges from quantum mechanics under coarsening of measurement times using the quantum violation of macrorealism as a tool, the present paper can also be useful in the context of implementing quantum information theoretic tasks based on temporal correlations. This is because non-classicality of temporal correlations has been employed in information theoretic
tasks like quantum computation \cite{app1}, randomness generation \cite{app2} and secure key distribution \cite{app3}. It is thus important to study how such non-classicality of temporal correlations persists in the real experimental scenario by taking into account the various types of coarsening of measurements.

\section*{Acknowledgement} 

D. D. acknowledges the financial support from University Grants Commission (UGC), Government of India. D. H. acknowledges the support from National Academy of Sciences, India (NASI).

\end{document}